\documentclass[12pt]{JHEP3}
\usepackage{amsmath,amssymb,epsfig,graphicx}

\newcommand{\be}{\begin{equation}}
\newcommand{\ee}{\end{equation}}
\newcommand{\beqa}{\begin{eqnarray}}
\newcommand{\eeqa}{\end{eqnarray}}

\def\eeq{\end{equation}}

\relax

\title{Medium-evolved fragmentation functions}

\author{N\'estor Armesto${}^{1}$, Leticia Cunqueiro${}^{1}$, Carlos A.
Salgado${}^{1,2}$
and
Wen-Chang Xiang${}^{3}$\footnote{Present address:
Fakult\"at f\"ur Physik, Universit\"at Bielefeld, D-33501
Bielefeld, Germany.}
\\
\vspace{0.1in}

${}^{1}$ Departamento de F\'\i sica de Part\'\i culas and IGFAE,
Universidade de Santiago de Compostela, E-15782 Santiago de Compostela, Spain
\vspace{0.1in}

${}^{2}$ Dipartimento di Fisica, Universit\`a di Roma ``La Sapienza"
and INFN, Roma, Italy
\vspace{0.1in}

${}^{3}$ Institute of Particle Physics, Huazhong Normal University, Wuhan 430079, China

\vspace{0.1in}

E-mail addresses: {\tt nestor@fpaxp1.usc.es, leticia@fpaxp1.usc.es, carlos.salgado@cern.ch,
wcxiang@physik.uni-bielefeld.de}
}

\abstract{Medium-induced gluon radiation is usually identified as the dominant dynamical mechanism underling the {\it jet quenching} phenomenon observed in heavy-ion collisions. In its actual implementation, multiple medium-induced gluon emissions are assumed to be independent, leading, in the eikonal approximation, to a Poisson distribution. Here, we introduce a medium term in the splitting probabilities so that both medium and vacuum contributions are included on the same footing in a DGLAP approach. The improvements include energy-momentum conservation at each individual splitting, medium-modified virtuality evolution and a coherent implementation of vacuum and medium splitting probabilities. Noticeably, the usual formalism is recovered when the virtuality and the energy of the parton are very large. This leads to a similar description of the suppression observed in heavy-ion collisions with values of the transport coefficient of the same order as those obtained using the {\it quenching weights}.}

\preprint{
January 2008\\
Roma-1-1460/2007}

\begin{document}


\section{Introduction}
\label{intro}

Hard processes are ideal tools to characterize the medium produced in heavy-ion collisions \cite{yr, Salgado:2007rs}. At large enough virtuality the perturbative partonic cross section is unchanged and all medium effects appear as modifications of the long distance parton distributions (PDF) and fragmentation functions (FF). An excellent example is the production of particles at high transverse momentum in which large modifications of the fragmentation functions have been experimentally observed \cite{RHIC}.

In the vacuum, the fragmentation of a perturbatively produced parton is well described by re-summing the leading soft and collinear singularities leading to an ordered shower structure described by the DGLAP evolution equations \cite{dglap}. 
In heavy-ion collisions, an enhanced splitting probability is expected due to the additional medium-induced radiation by gluon bremsstrahlung off the fast parton, producing a softening of the associated jet structures. This effect is commonly identified as the dominant dynamical mechanism producing the strong suppression of high-$p_T$ hadrons observed experimentally at RHIC \cite{RHIC}. Provided the geometry of the collision is known, the only parameter controlling this suppression is the transport coefficient, $\hat q$, characterizing the average squared transverse momentum acquired by the emitted gluon per mean free path in the medium. Hence, a one-to-one correspondence between the energy loss and the jet broadening exists as both are controlled by the same parameter \cite{Baier:1996sk}.

The above properties are based on the one-gluon inclusive spectrum computed in Refs. \cite{Baier:1996sk,Zakharov:1997uu,Wiedemann:2000za,Gyulassy:2000er} (see also
related approaches in \cite{Wang:2001if, Arnold:2002ja}). For practical applications, however, more differential quantities, including exclusive one, two,$\dots$ gluon emissions are needed. 
A successful phenomenology, based on this radiative energy loss, assumes a separation between vacuum and medium contributions to the shower development in which the "medium" radiation occurs before in time than the "vacuum" radiation, the latter taking place after the fast parton exits the medium. In this way, an independent emission approximation is taken for the first \cite{Salgado:2003gb,quenwei,Gyulassy:2001nm,Jeon:2003gi} while the usual (vacuum) DGLAP evolution describes the second. In the eikonal approximation for multiple gluon radiation, the energy loss distributions are Poissonian and normally known as {\it quenching weights} \cite{Salgado:2003gb,quenwei}. One caveat of this formalism is the different role of the medium and vacuum radiations which looks artificial. Further limitations are the treatment of energy constrains and the role of virtuality. In the present paper we study these limitations and propose a new formalism in which the medium corrections are systematically included as an additive term in the splitting functions. This new term is directly taken from the known medium-induced gluon radiation spectrum by comparing the leading vacuum contributions. Our main assumption is the existence of an ordering variable also for the medium, which we will take to be the virtuality. This assumption is in agreement with the findings in  \cite{Wang:2001if}  and has been used before in \cite{Polosa:2006hb,Borghini:2005em,Sapeta:2007ad}. Interestingly, we show that when the virtuality and the kinematic constrains coincide, the usual formalism, which provides an excellent agreement with experimental data on inclusive high-$p_T$ particle production, is recovered.

Our more refined treatment of the fragmentation functions is mainly motivated by the imminence of the LHC heavy-ion program, in which real jets will be measured in heavy-ion collisions for the first time and their fragmentation functions reconstructed \cite{Carminati:2004fp, Alessandro:2006yt, D'Enterria:2007xr,atlas}. We present full calculations of medium-modified fragmentation functions at different energies and virtualities easily reachable within the first years of the program. Although our overall aim is the LHC physics, our work will also find applications to RHIC physics. In particular, we recalculate the high-$p_T$ suppression of light hadrons and check that the results agree with the ones already obtained within the Poisson approximation using the  quenching weights.

The paper is organized as follows: In Section \ref{sec2} we first present the medium-modified splitting functions, which is then used to compute the Sudakov form factors and DGLAP evolution. In Section \ref{sec3} we present our results for the medium-modified fragmentation functions and calculate the inclusive particle suppression to check, in Section \ref{sec3-3}, the degree of agreement with the previous formalism. Finally we present our conclusions. In Appendix \ref{app1} a derivation of the quenching weights is obtained as a limiting case of our formalism for large fraction of momentum and virtuality.

\section{Medium-evolved fragmentation functions}
\label{sec2}

The DGLAP evolution equations for the fragmentation function $D(x,t)$ reads, omitting parton type indices,
\begin{equation}
t\frac{\partial}{\partial t} D(x,t)=\int_x^1\frac{dz}{z}\frac{\alpha_s}{2\pi}\ P(z)\, D\left(\frac{x}{z},t\right).
\label{eq:DGLAP}
\end{equation}
Here, $P(z)$ is the splitting function describing the branching of a parton into two new ones with fractions of momenta $z$ and $1-z$. For our implementation of medium effects in the fragmentation functions we recall the  probabilistic interpretation of the (LO) DGLAP evolution which is evident from its integral formulation (see e.g. \cite{ellis})
\begin{equation}
D(x,t)=\Delta(t)D(x,t_0)+\Delta(t)\int_{t_0}^t \frac{dt_1}{t_1}
\frac{1}{\Delta(t_1)} \int \frac{dz}{z} \, P(z)
D\left(\frac{x}{z},t_1\right).
\label{eq:dglapsud}
\end{equation}
The first term on the right-hand side in this expression corresponds  to the contribution with no splittings between $t_0$ and $t$ while the second one gives the evolution when some finite amount of radiation is present. The evolution is controlled by the Sudakov form factors
\begin{equation}
\Delta(t)=\exp{\left[-\int_{t_0}^{t} {dt^\prime \over t^\prime}
\int dz {\alpha_s(t^\prime,z)
\over 2 \pi} P(z,t^\prime)\right]},
\label{eq:sudakovs}
\end{equation}
with the interpretation of the probability of no resolvable branching between the two scales $t$ and $t_0$. 

The definition of the Sudakov form factors and its probabilistic
interpretation depend on the cancellation of the different divergencies
appearing in the corresponding Feynman diagrams, see e.g.
\cite{Dokshitzer:1978hw}. Although such cancellation has never been proved on general grounds for partons re-scattering in a medium,  it has been found in \cite{Wang:2001if} that, under certain assumptions, all the medium effects can be included in a redefinition of the splitting function
\begin{equation}
P^{\rm tot}(z)= P^{\rm vac}(z)+\Delta P(z,t),
\label{eq:medsplit}
\end{equation}
where we have labeled as "vac" the corresponding vacuum splitting function. The main assumption to arrive at (\ref{eq:medsplit}) is the independence of the multiple gluon emission when re-scattering with the nuclei is present so that an exponentiation of the splittings is possible. This independence could be broken if non-trivial color reconnections are present 
\cite{JalilianMarian:2004da,Baier:2005dv} but the size of this corrections for the DGLAP kinematics has not been computed. In this exploratory study, we will assume that the definition (\ref{eq:medsplit}) can be extended to the medium case. Similar assumptions have been done in \cite{Polosa:2006hb,Borghini:2005em,Sapeta:2007ad} using simplified forms of the medium term $\Delta P$ which we will now improve by taking the full splitting probability as given by the medium-induced gluon radiation spectrum. As we will show, this procedure recovers the usually employed quenching weights \cite{Salgado:2003gb,quenwei}, making a connection with previous phenomenology in the field.

We now describe the steps leading from the single inclusive
distribution of gluons radiated in a medium from a parent parton, to the
medium-modified fragmentation functions. We start by defining our
medium-modified splitting functions, then we compute the corresponding
Sudakov form factors, to finally write the modified DGLAP-like evolution
equations that we use, and the initial conditions for such evolution.

\subsection{Medium-modified splitting functions}
\label{sec2-1}

The formalism of medium-induced gluon radiation
\cite{Baier:1996sk,Zakharov:1997uu,Wiedemann:2000za,Gyulassy:2000er}
provides the single inclusive distribution of gluons emitted with energy
$\omega$ and transverse momentum ${\bf k}_\perp$, by
medium-induced radiation from a parent parton with energy $E$, $x=\omega/E$.
For the case of a massless
parent the general formula for the emitted energy reads \cite{Wiedemann:2000za}
\begin{eqnarray}
  \omega\frac{dI}{d\omega\, d{\bf k}_\perp}
  &=& {\alpha_s\,  C_R\over (2\pi)^2\, \omega^2}\,
    2{\rm Re} \int_{0}^{\infty}\hspace{-0.3cm} dy_l
  \int_{y_l}^{\infty} \hspace{-0.3cm} d\bar{y}_l
   \int d{\bf u}\, e^{-i{\bf k}_\perp \cdot {\bf u}}   \,
  e^{ -\frac{1}{2} \int_{\bar{y}_l}^{\infty} d\xi\, n(\xi)\,
    \sigma({\bf u}) }\,
  \nonumber \\
  & \times& {\partial \over \partial {\bf y}}\cdot
  {\partial \over \partial {\bf u}}\,
  \int_{{\bf y}=0={\bf r}(y_l)}^{{\bf u}={\bf r}(\bar{y}_l)}
  \hspace{-0.5cm} {\cal D}{\bf r}
   \exp\left[ i \int_{y_l}^{\bar{y}_l} \hspace{-0.2cm} d\xi
        \frac{\omega}{2} \left(\dot{\bf r}^2
          - \frac{n(\xi) \sigma\left({\bf r}\right)}{i\, \omega} \right)
                      \right]\, .
    \label{2.1}
\end{eqnarray}
Here, the Casimir factor
$C_R = \frac{4}{3}$ for a parent quark and 3 for a parent gluon,
determines the coupling strength of the emitted gluon to the parent.
Eq.~(\ref{2.1}) re-sums the effects
of arbitrary many medium-induced scatterings to leading order in $1/E$.
Properties of the medium enter (\ref{2.1}) via the product
of the time-dependent density $n(\xi)$ of scattering centers times
the strength of a single elastic scattering $\sigma({\bf r})$.
A detailed discussion of
(\ref{2.1}) including the physical interpretation of the
integration variables ($y_l$, $\bar{y}_l$, ${\bf y}$, ${\bf u}$,
$\xi$) can be found in Refs.~\cite{Wiedemann:2000za,Salgado:2003gb}
for the massless case and
in \cite{Armesto:2003jh} for the massive one.

In (\ref{2.1}) the integration in $y_l$ ($\bar y_l$) signals the longitudinal position of the gluon radiation vertex in the amplitude (complex conjugate amplitude). For the case of high-$p_t$ particle production, the parent parton is produced inside the medium and three cases can be distinguished depending on the position of the radiation vertex begin inside or outside the medium in both amplitude and complex conjugate amplitude and the interference of the two. In the case that both radiation vertex are outside the medium the usual vacuum radiation spectrum is obtained. This spectrum is normally subtracted from the total to define the medium-induced gluon radiation $I^{\rm med}$ as
\begin{eqnarray}
  \omega\frac{dI}{d\omega\, d{\bf k}_\perp}
  = \omega\frac{dI^{\rm vac}}{d\omega\, d{\bf k}_\perp}
    + \omega\frac{dI^{\rm med}}{d\omega\, d{\bf k}_\perp}\, .
  \label{2.4}
\end{eqnarray}
The vacuum spectrum defines the (vacuum) splitting function, which at small fraction $x=1-z$ is
%
\begin{equation}
\frac{dI^{\rm vac}}{dz\, d{\bf k}_\perp^2}=\frac{\alpha_s}{2 \pi}
\frac{1}{{\rm k}_\perp^2} P^{\rm vac}(z),\ \ P^{\rm vac}(z)
\simeq \frac{2 C_R}{1-z}\,.
\label{vacsplit}
\end{equation}

For the medium part, two different approximations are usually employed in (\ref{2.1}), the
exact solution of the path integral being unknown. The first one corresponds
to expanding (\ref{2.1}) in powers of the opacity $n(\xi) \sigma({\bf r})$
\cite{Gyulassy:2000er,Wiedemann:2000tf}. The second one, which we will follow
in this work\footnote{Detailed comparisons of both approximations in the
massless and massive cases can be found in \cite{Salgado:2003gb} and
\cite{Armesto:2003jh} respectively.},
corresponds to
\begin{eqnarray}
  n(\xi)\, \sigma({\bf r}) \simeq \frac{1}{2}\, \hat{q}(\xi)\, {\bf r}^2\, ,
  \label{4.1}
\end{eqnarray}
where $\hat{q}(\xi)$ is the transport coefficient \cite{Baier:1996sk}
which characterizes the medium-induced transverse momentum squared
$\langle q_\perp^2\rangle$ transferred to the projectile per unit
path length $\lambda$. Provided the collision geometry is known, all medium properties are encoded in the transport coefficient. In the approximation (\ref{4.1}), the path
integral in (\ref{2.1}) is equivalent to that of a harmonic oscillator and its solution is known.
The resulting expressions can be found in
\cite{Wiedemann:2000za,Salgado:2003gb}.

For a static medium\footnote{The case of a medium which expands and gets
dilute can be mapped onto a static one with a redefinition of the transport
coefficient \cite{Baier:1998yf, Gyulassy:2000gk, Salgado:2002cd}.},
the spectrum of medium-induced radiated gluons is a function of two dimensionless variables $\omega/\omega_c$ and $\kappa^2$ defined by
\begin{equation}
\omega_c=\frac{1}{2} \hat{q} L^2,\ \ \kappa^2=\frac{{\bf k}_\perp^2}
{\hat{q}L}\,.
\label{scalvar}
\end{equation}
The  shape of this function is shown in Fig. \ref{fig1} for different values of the energy of the radiated gluon.

Direct comparison with Eq. (\ref{vacsplit}) defines the medium-induced part of the splitting probability
\begin{equation}
\Delta P(z,t)\simeq \frac{2 \pi  t}{\alpha_s}\, 
\frac{dI^{\rm med}}{dzdt} ,
\label{medsplit}
\end{equation}
where $\omega=(1-z) E$ and ${\bf k}_\perp^2=z(1-z)t$ are taken in the medium-induced gluon radiation spectrum - parton masses have been neglected.

\FIGURE[h]{\epsfig{file=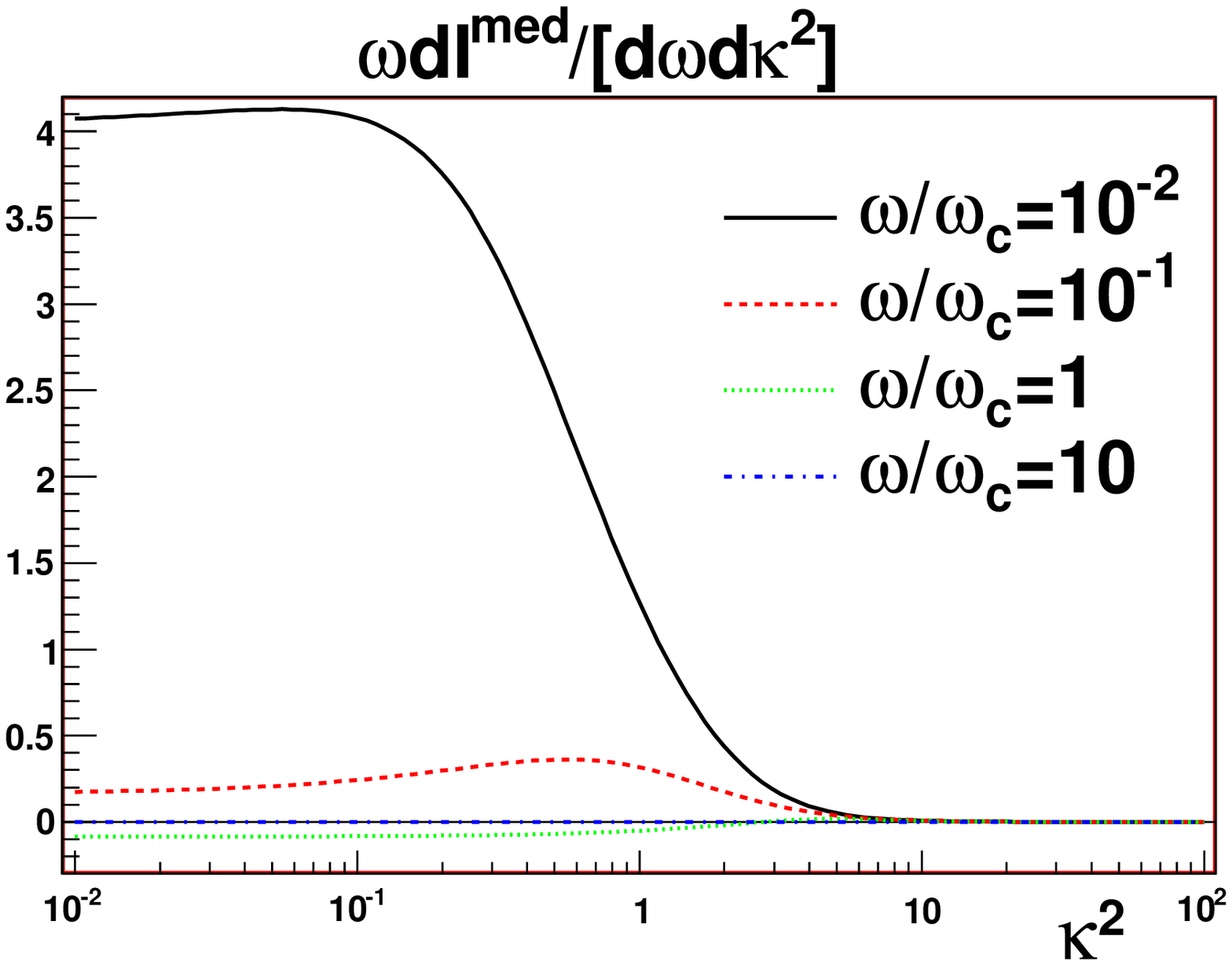,width=11cm}
\caption{Single-gluon inclusive spectra versus $\kappa^2$ for $\alpha_s=1=C_R$ and 
different values of $\omega/\omega_c$.}
\label{fig1}}

At this level, $g\to gg$ and $q\to qg$ splitting functions are different by
simply the Casimir color factors.
Let us now discuss finite-$x$ corrections to the leading behavior described in the expressions above. The helicity averaged contribution for the case when the radiation vertices (in amplitude and complex conjugate amplitude) take place outside the medium recovers the usual vacuum splitting functions\footnote{The reason for this is that in the eikonal approximation, the only effect of the medium on the parent parton  before the splitting is a color rotation which cancels in the medium averages \cite{Baier:1996sk,Zakharov:1997uu,Wiedemann:2000za} -- see also \cite{CasalderreySolana:2007pr} for a recent discussion.} and the full lowest order form will be used for them. In the medium case, the helicity averaged contributions to the radiation vertex factorize as a multiplicative factor of the form $(1-z)P^{\rm vac}(z)/(2C_R)=(1+z^2)/2$ for the case where the parent parton is a quark  \cite{Baier:1996sk,Zakharov:1997uu,Wiedemann:2000za,Gyulassy:2000er} and similarly for the gluon. So, in summary, to take into account the finite-$x$ correction to the medium splitting function, we multiply the right-hand side of (\ref{medsplit}) by $(1+z^2)/2$ if the parent parton is a quark and by $z$ if the parent parton is a gluon. In the second case we impose a symmetrization around $z=1/2$.
We have also checked that making the substitution  $\omega\to z(1-z) E$ in (\ref{medsplit}) \cite{Zakharov:1997uu} produces only small corrections.
The impact of these large $x$ modifications is small as we show
numerically at the level of the Sudakov form factors in the next Subsection.

For the coupling constant $\alpha_s(Q^2)$,
we will use a lowest-order running with
3 quark flavors, with $\Lambda_{QCD}=0.236$ GeV which gives $\alpha_s(m_Z)=0.1172$. In the infrared, we freeze the coupling at $Q^2=4$ GeV$^2$. The
scale is taken to be $Q^2=z(1-z)t$ which corresponds to the squared transverse
momentum of the emission\footnote{The medium-induced gluon spectrum has been
deduced for a fixed coupling constant. Nevertheless, arguments have been given
\cite{Baier:1996sk}
that the transverse momentum of the emitted gluon provides the right scale in
this problem.}. This choice corresponds to the implementation of
angular ordering in the emissions, see e.g. \cite{ellis}. We have checked that all qualitative conclusions we extract with this implementation hold when we implement a different scale for the running of the coupling (e.g. $Q^2=t$) or when we freeze the coupling to a larger value (e.g. at $Q^2=2$ GeV$^2$).

\subsection{Sudakov form factors and medium-modified fragmentation functions }
\label{sec2-2}

The Sudakov form factor, including summation of different parton species and the relevant kinematical limits reads  \cite{ellis},
\begin{equation}
\Delta_i(t)=\exp{\left[-\int_{t_0}^{t} {dt^\prime \over t^\prime}
\int_{z_{min}(t^\prime)}^{1-z_{min}(t^\prime)} dz {\alpha_s(t^\prime,z)
\over 2 \pi} \sum_{j}P_{i\to j}(z,t^\prime)\right]},
\label{sudakovs}
\end{equation}
with $z_{min}(t)=t_0/t$. Here, $t_0$ is some scale below which non-perturbative effects (i.e. hadronization)
become important. We take
$t_0=1$ GeV$^2$. Note that the evolution only makes sense from $t$
up to $2t_0$, to prevent $z_{min}(t) > 1-z_{min}(t)$. At this level, we
consider $t_0$ to be independent of the parent or daughters being gluons or
any quark flavors. In (\ref{sudakovs}) the summation over flavors ($j=g,\, u,\, \bar u,\, d,\dots$) is made explicit to indicate that all possibilities $q\to qg$, $g\to gg$ and $g\to q\bar q$ are taken into account in the evolution. For the case $g\to q\bar q$, the vacuum splitting function is always taken, so it acquires no medium modification as it is subleading in energy; numerically, this contribution is anyway very small\footnote{Notice that when the evolution takes place in the medium, there is a chance that the parent parton {\it fuses} with a parton from the medium, $gg\to g$, $qg\to q$, $q\bar q\to g$, and conversions of a gluon jet into a quark one or viceversa may take place. This possibility has been taken into account in e.g. \cite{Jeon:2003gi,Ko:2007zza}. The effect of these corrections, not included in the present work, is not large and restricted to the small- and moderate-$p_T$ part of the spectrum.}.

Using the medium-modified splitting functions defined by (\ref{eq:medsplit}), 
we compute the corresponding Sudakov form factors for quarks and gluons. In
Fig. \ref{fig2} we show an example for several values of medium parameters
$\hat{q}$ and $L$, and for parton energies $E=10$ and $100$ GeV.
The influence of the medium, and the difference for quarks and gluons,
are clearly visible (and the relation of the
splitting functions at small $x$ given by the
Casimir factors almost exactly fulfilled). We also demonstrate the small
effect produced by the large-$x$ corrections that we introduce.
\FIGURE[h]{\epsfig{file=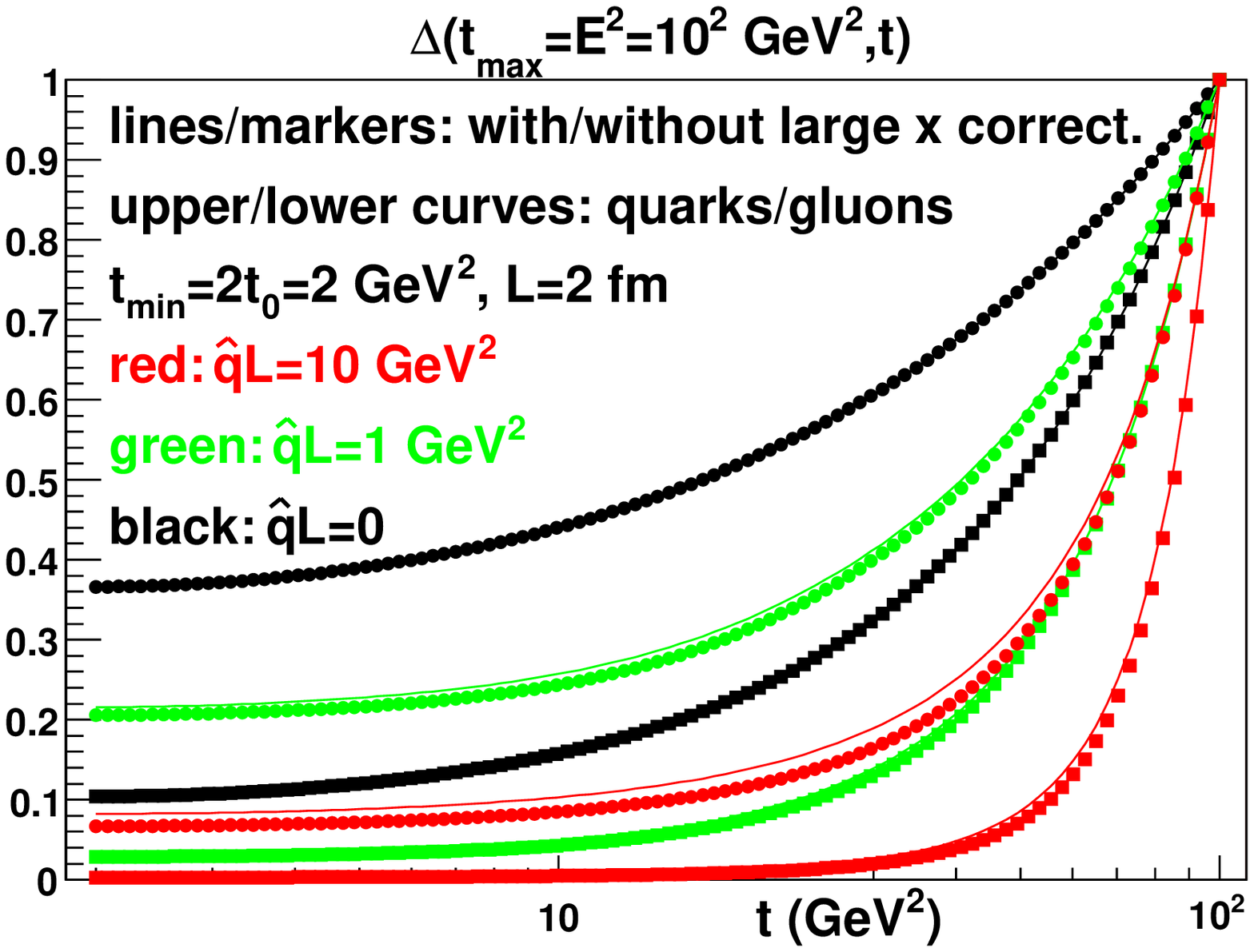,width=8.85cm}\epsfig{file=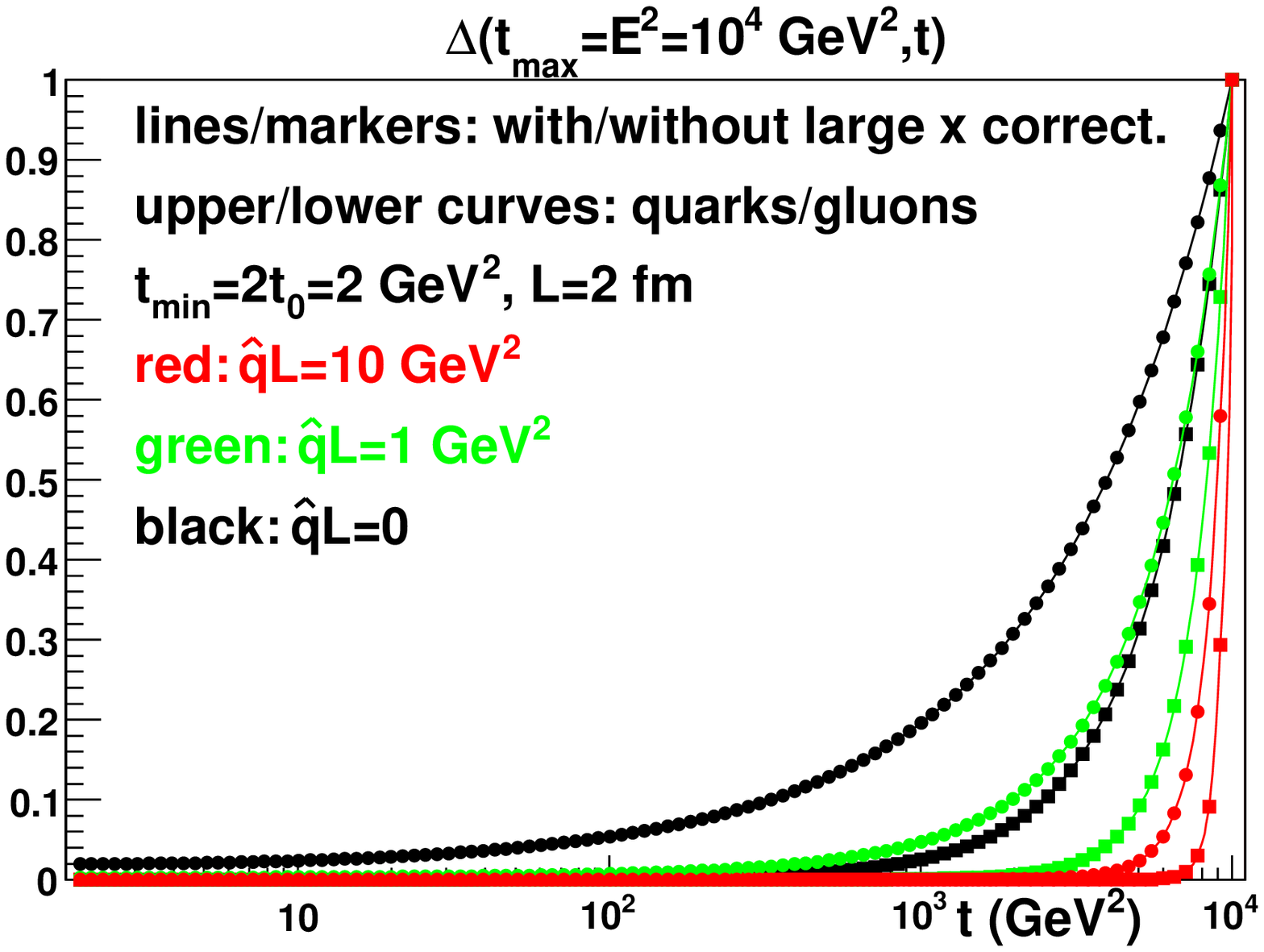,width=8.85cm}
\epsfig{file=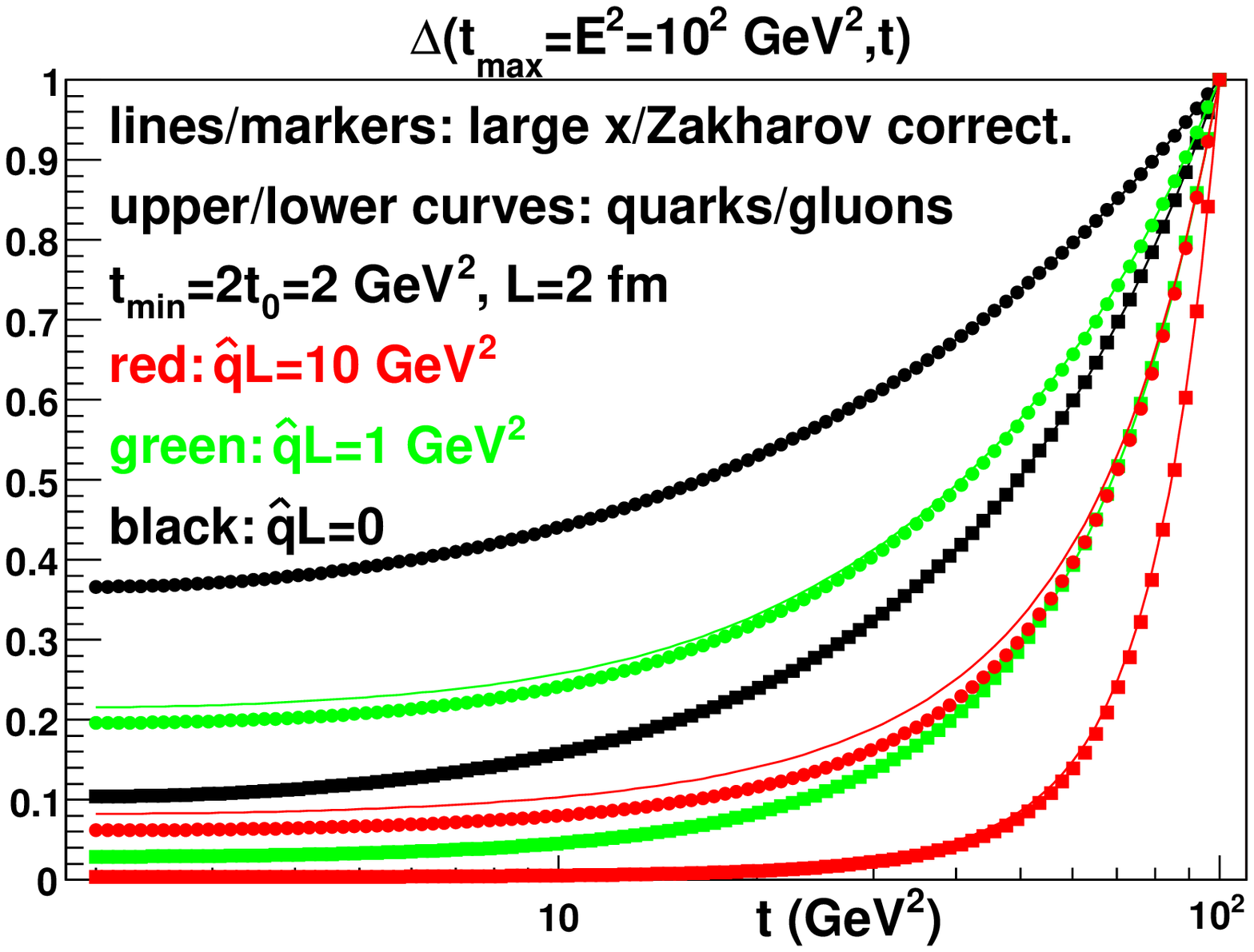,width=8.85cm}\epsfig{file=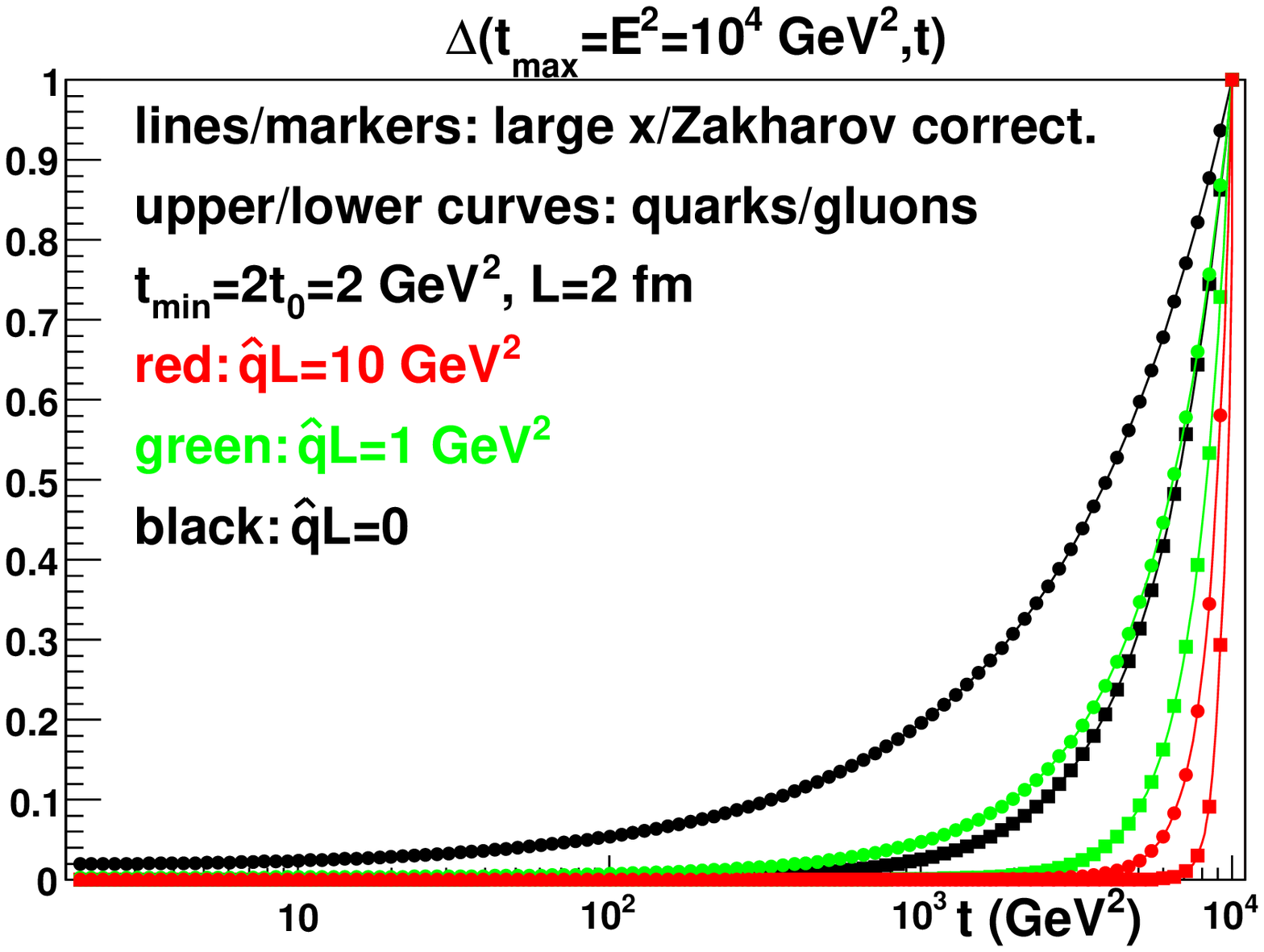,width=8.85cm}
\caption{Sudakov forms factors for quarks (upper curves and symbols in each
plot) and
gluons (lower curves and symbols in each plot),
for parton energies $E=10$ (plots on the
left) and 100 (plots on the right) GeV, and for different medium parameters and
large-$x$ corrections to the medium splitting functions. In the legends on the
plots, 'large x' refers to the (default) two first corrections explained in
Subsection \ref{sec2-1}, and 'Zakharov corrections' to the third correction
explained in that Subsection.}
\label{fig2}}


In order to compute the medium-modified fragmentation functions, the only missing ingredient is the initial condition for the DGLAP evolution (\ref{eq:DGLAP}). This initial condition is a non-perturbative quantity describing the formation of the final hadron. Although a medium-modification of this quantity could be possible, we will simply assume that this initial condition is unchanged and equal to the vacuum one

\begin{equation}
D^{\rm med}(x,t_0)=D^{\rm vac}(x,t_0)\, ,
\label{eq:initial}
\end{equation}
and we take $D^{\rm vac}(x,t_0)$ from \cite{Kniehl:2000fe}. The motivation for this ansatz is the following: in hadronic collisions, particles produced at high enough transverse momentum hadronize outside the medium. Eq. (\ref{eq:initial}) assumes that this non-perturbative hadronization is not modified by the medium, whose effect is only to modify the perturbative associated radiation\footnote{For processes with different kinematic conditions  \cite{Airapetian:2007vu} this assumption could not hold, but the negligible effects seen in dAu data at RHIC \cite{RHIC} indicate that this is a reasonable assumption for particle production at high-p$_t$ in nuclear collisions.}. All present radiative energy loss formalisms rely on this assumption (see the Appendix \ref{app1} for the equivalence with these formalisms).

The medium-modified evolution given by Eqs. (\ref{eq:dglapsud}), (\ref{eq:sudakovs}) and (\ref{eq:medsplit}) and the initial condition (\ref{eq:initial}) complete the formulation of our approach.

\section{Results}
\label{sec3}


\label{sec3-1}
We solve Eq. (\ref{eq:dglapsud}) numerically in order to obtain our medium-modified fragmentation functions (MMFF). We use a brute-force method with a 4th-order adaptative Runge-Kutta for the evolution in virtuality plus a Gaussian quadrature for the integration in momentum fraction (and virtuality for the Sudakov form factors).
For the initial conditions we use the Kniehl-Kramer-Potter (KKP) \cite{Kniehl:2000fe}
parameterizations at $Q^{2}=2$ GeV$^{2}$ for pions and we consider only three flavors u,
d, s\footnote{We have checked that the use of a different parametrization \cite{de Florian:2007hc} for the initial conditions does not change the conclusions of our study.}. We have checked that our simplified evolution reproduces the KKP results better than 40 \% in the region of validity of the KKP parametrizations ($0.1<z<0.9$) up to the highest virtualities we have considered. 

In Figs. \ref{fig311} and \ref{fig311p} we show our results for the medium-evolved
fragmentation functions onto pions for two different parton energies ($10$ GeV, relevant
for RHIC and $100$ GeV, relevant for the LHC), for different medium parameters and
different parton types. The main effect found is a suppression of the fragmentation functions at large
$z$ values and a corresponding enhancement at small $z$. 
These two features grow with increasing scale (the higher the scale, the higher
the length of evolution thus the greater the medium modifications).
They also increase with the transport coefficient and, as shown in
Fig. \ref{fig312},  with the medium length.
These medium effects decrease however with the parent parton energy. The
qualitative reason for this is that the energy
loss becomes more and more energy-independent with increasing energy and thus the fractional energy loss decreases with increasing energy.
As expected, the medium modifications are greater for gluons than for quarks.

\FIGURE[h]{\epsfig{file=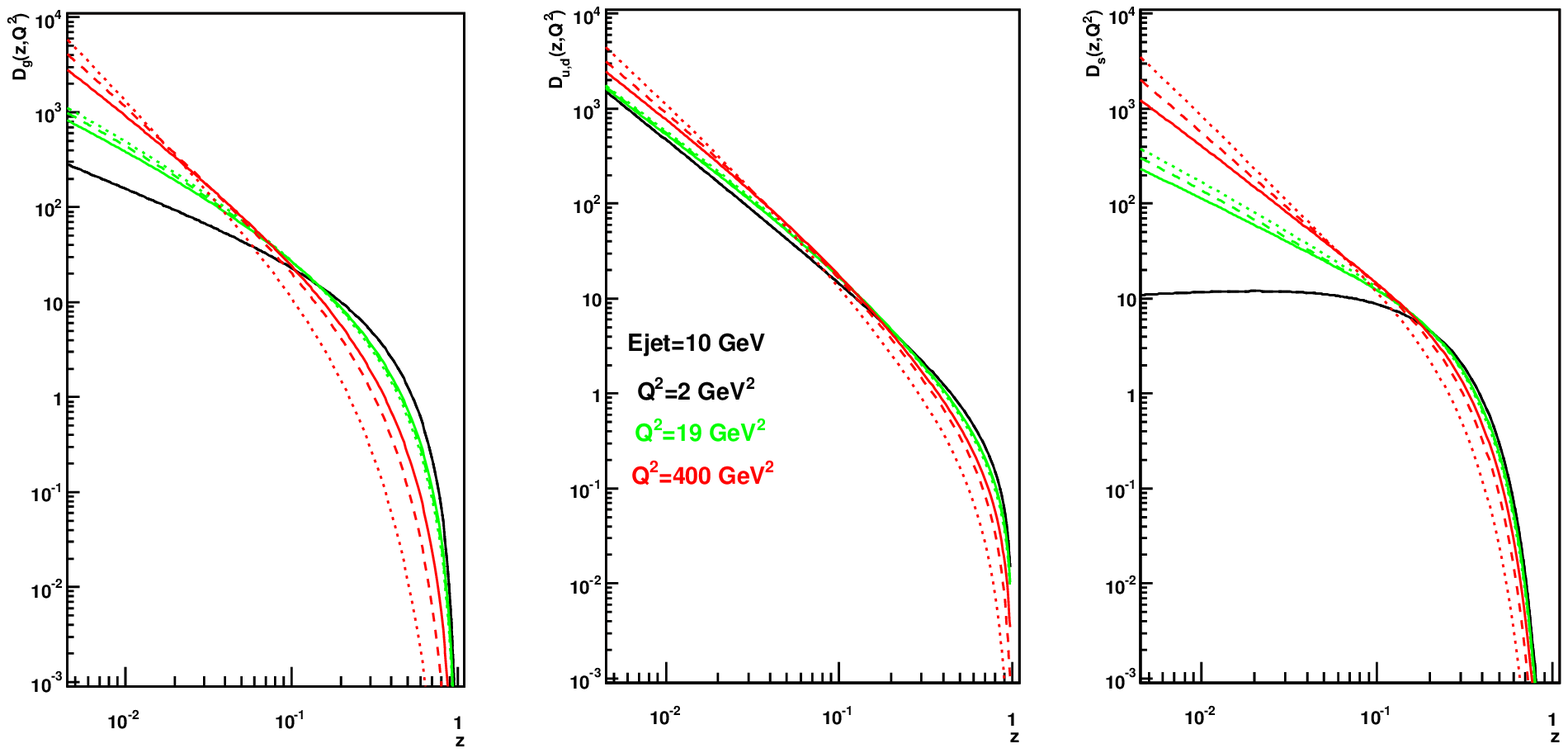,width=\textwidth}
\caption{Medium-modified fragmentation functions for gluons (left),
  u (or d) quarks (middle) and s quarks (right) as a function of the energy
  fraction carried by the hadron, $z$. The parent parton energy
  is $10$ GeV and the medium length is $L=2$ fm. For each of the three
  different scales ($Q^{2}=2$, $19$ and $400$ GeV$^{2}$ in black, green and
  red respectively) we plot the fragmentation functions at three different medium densities:
  vacuum (solid), $\hat{q}=1$ GeV$^{2}$/fm (dashed) and $\hat{q}=10$
  GeV$^{2}$/fm (dotted).}
\label{fig311}}

\FIGURE[h]{\epsfig{file=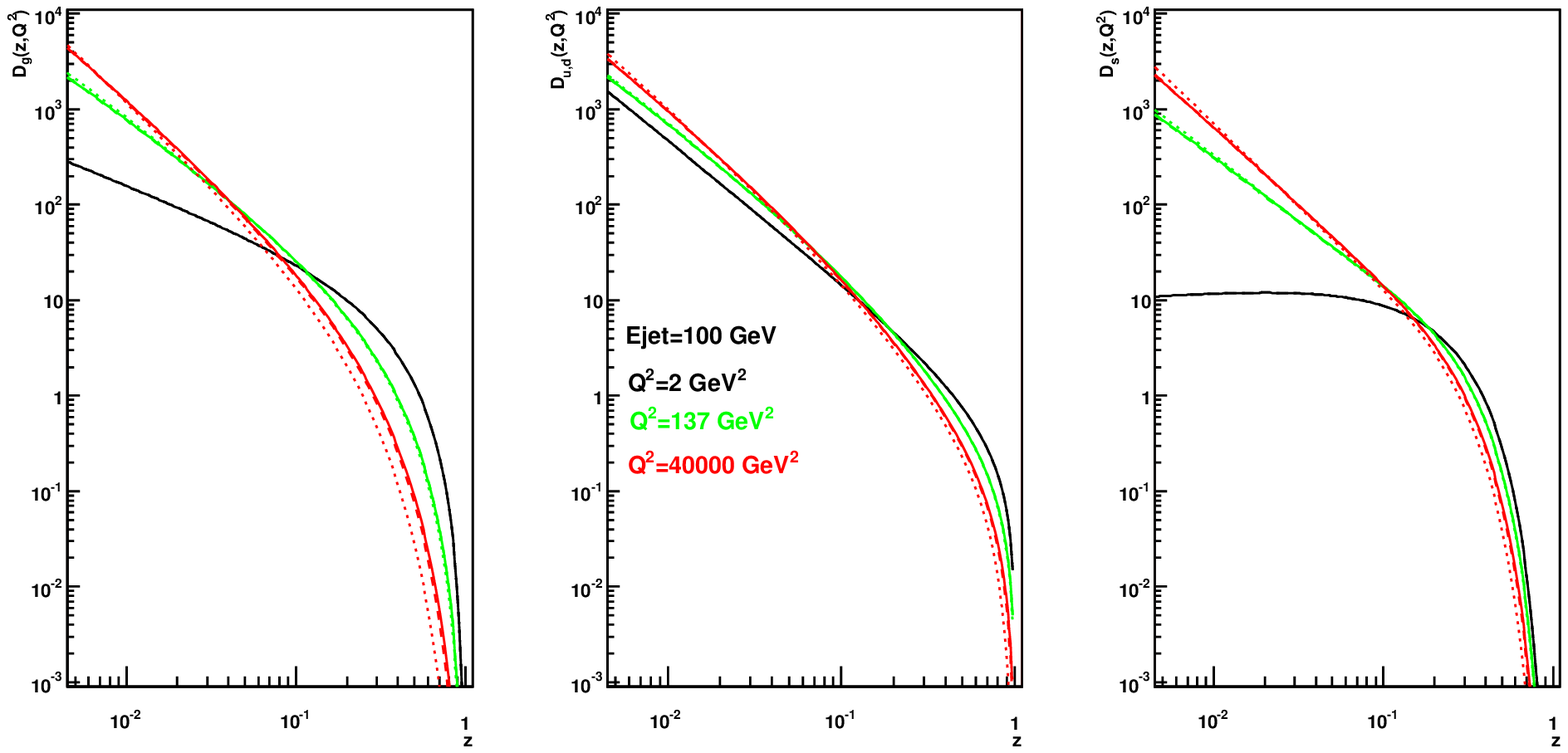,width=\textwidth}
\caption{The same as Fig. \ref{fig311} but for a parent parton energy
  of $E_{jet}=100$ GeV and scales $Q^{2}=2$, $137$ and $40000$ GeV$^{2}$ in
  black, green and red respectively.}
\label{fig311p}}

\FIGURE[h]{\epsfig{file=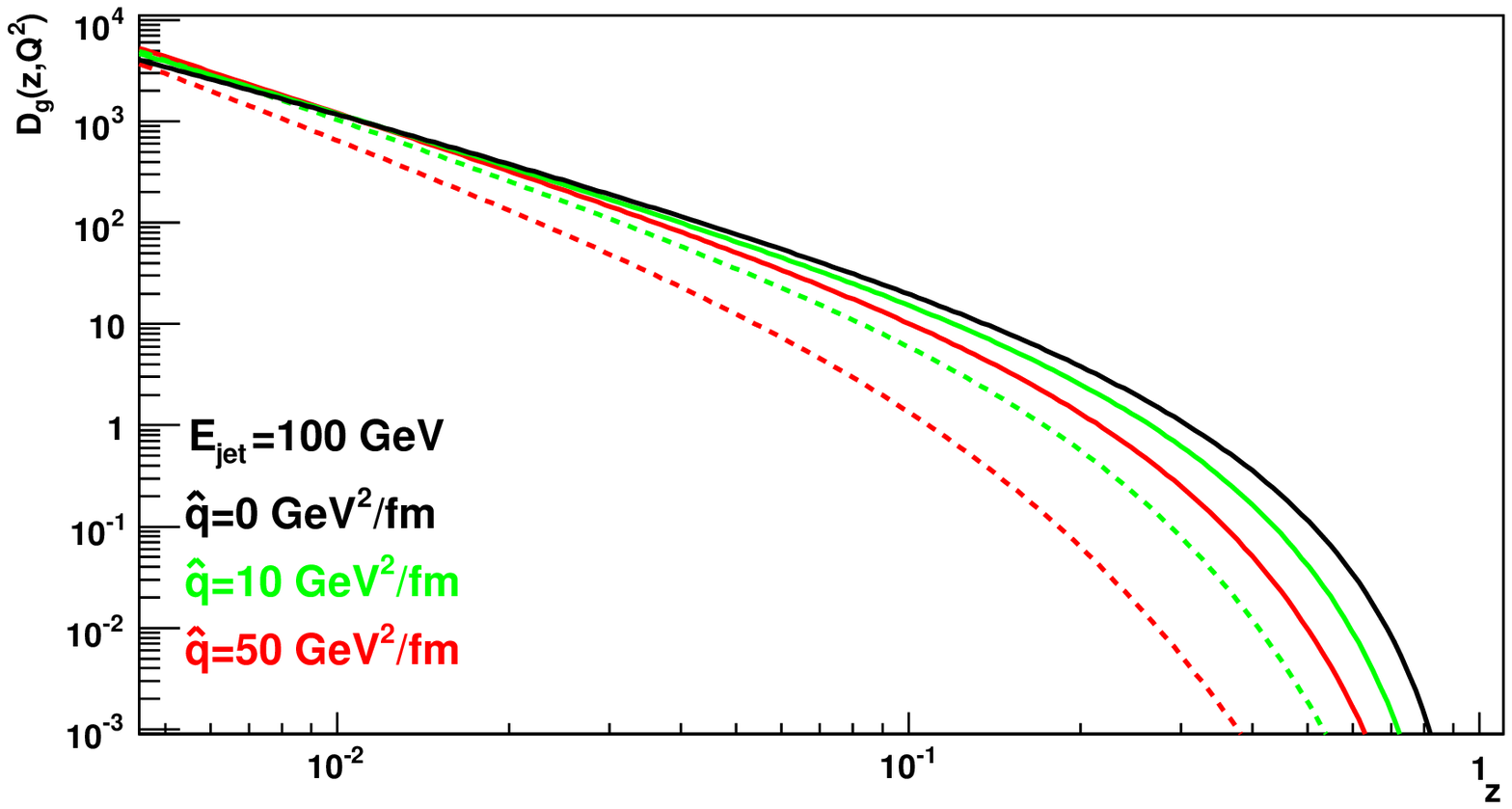,width=8.3cm,height=6.3cm}\epsfig{file=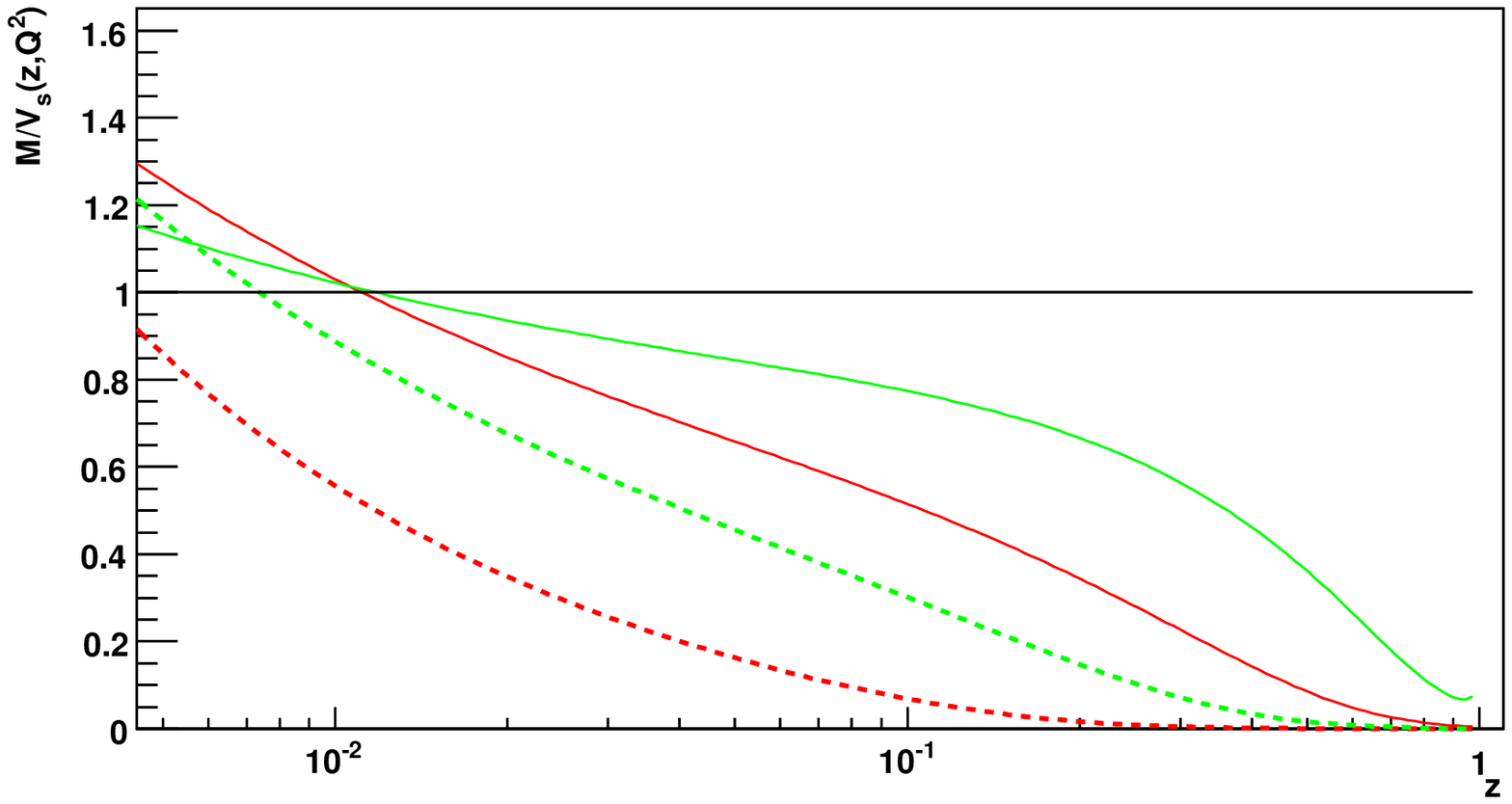,width=8.3cm,height=6.3cm}
 \caption{Left: Fragmentation function for gluons onto pions at $E_{jet}=100$
   GeV. Our results are plotted at $Q^{2}=E_{jet}^{2}$, for three different
  mediums: vacuum (black), $\hat{q}=10$ GeV$^{2}$/fm (green) and
  $\hat{q}=50$ GeV$^{2}$/fm (red) and for two different medium lengths:
  $2$ fm (solid) and $6$ fm (dashed). Right:  Medium to vacuum ratio of the gluon
  fragmentation functions for the same values as in the plot on the left.}
 \label{fig312}}

In order to check the feasibility of our approach and the compatibility of the extracted medium parameters with those obtained within other approaches, we compute the medium-modified spectra of neutral pions at RHIC. 
We standardly convolute our medium-modified fragmentation functions with the nuclear parton densities (we use the EKS98 parametrization from Ref. \cite{Eskola:1998df}) and the hard scattering elements to obtain the medium-modified particle
spectra following the factorization formula at LO:
\begin{equation}
\sigma^{AB \to h}=f_{A}(x_1,Q^{2})f_{B}(x_2,Q^{2})\otimes
\sigma(x_{1},x_{2},Q^{2}) \otimes D_{i \to h} (z,Q^{2}).
\label{eq:csec}
\end{equation}  
As it is common in the phenomenology of jet quenching in heavy ion collisions, we show our results for the nuclear modification factor defined as 
\begin{equation}
R_{AA}=\frac{\left.d\sigma^{AA}/d\eta dp_{T}^{2}\right|_{\eta=0}}
{\left.d\sigma^{pp}/ d\eta dp_{T}^{2}\right|_{\eta=0} }\,,
\label{eq:raa}
\end{equation}
where the numerator correspond to the cross section (\ref{eq:csec}) including the nuclear corrections to the PDFs and our MMFF and the denominator is computed with the proton PDFs and the vacuum fragmentation functions. 

\begin{figure}
\begin{minipage}{0.5\textwidth}
\includegraphics[width=\textwidth]{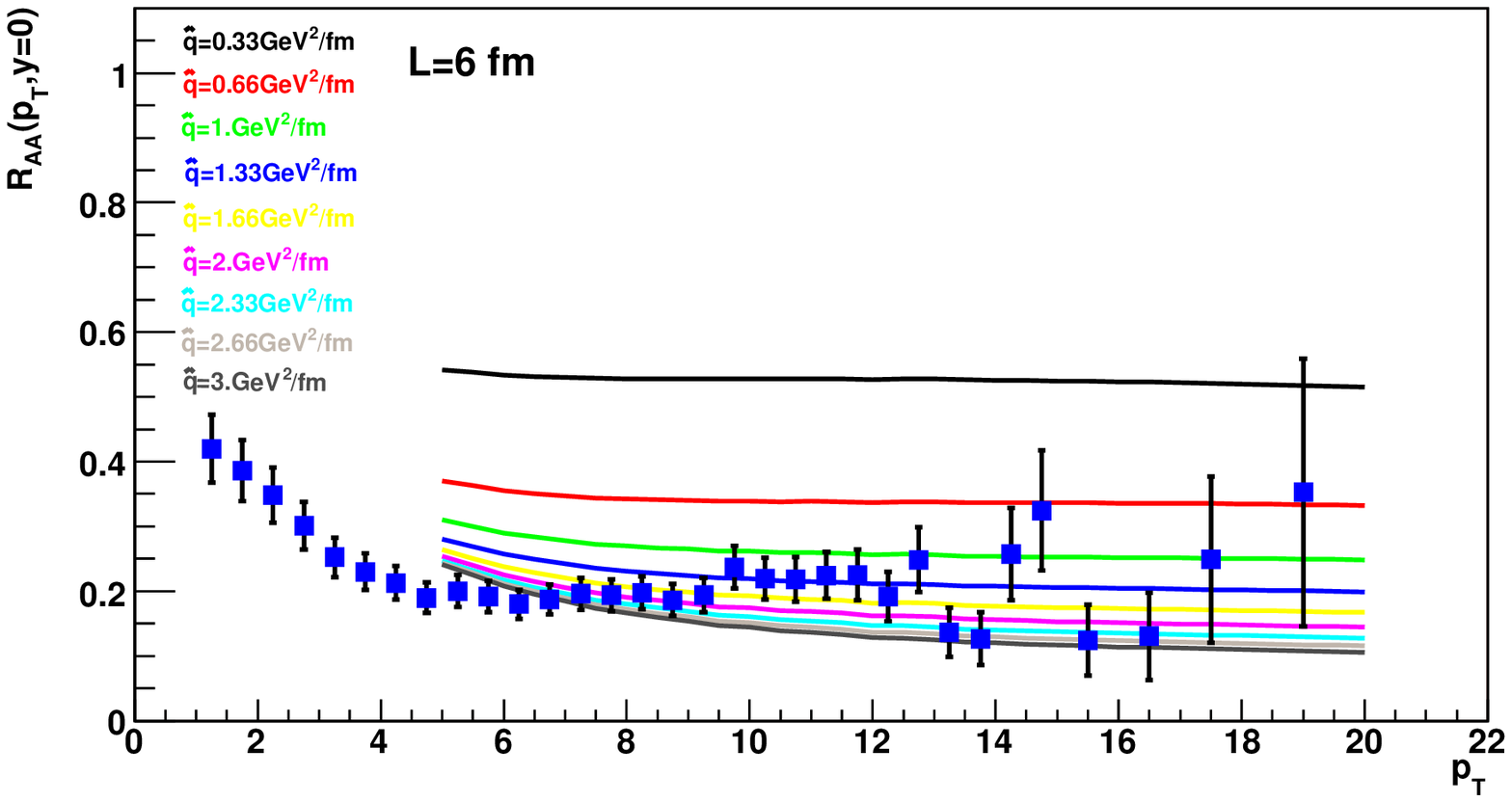}
\end{minipage}
\hfill
\begin{minipage}{0.5\textwidth}
\includegraphics[width=\textwidth]{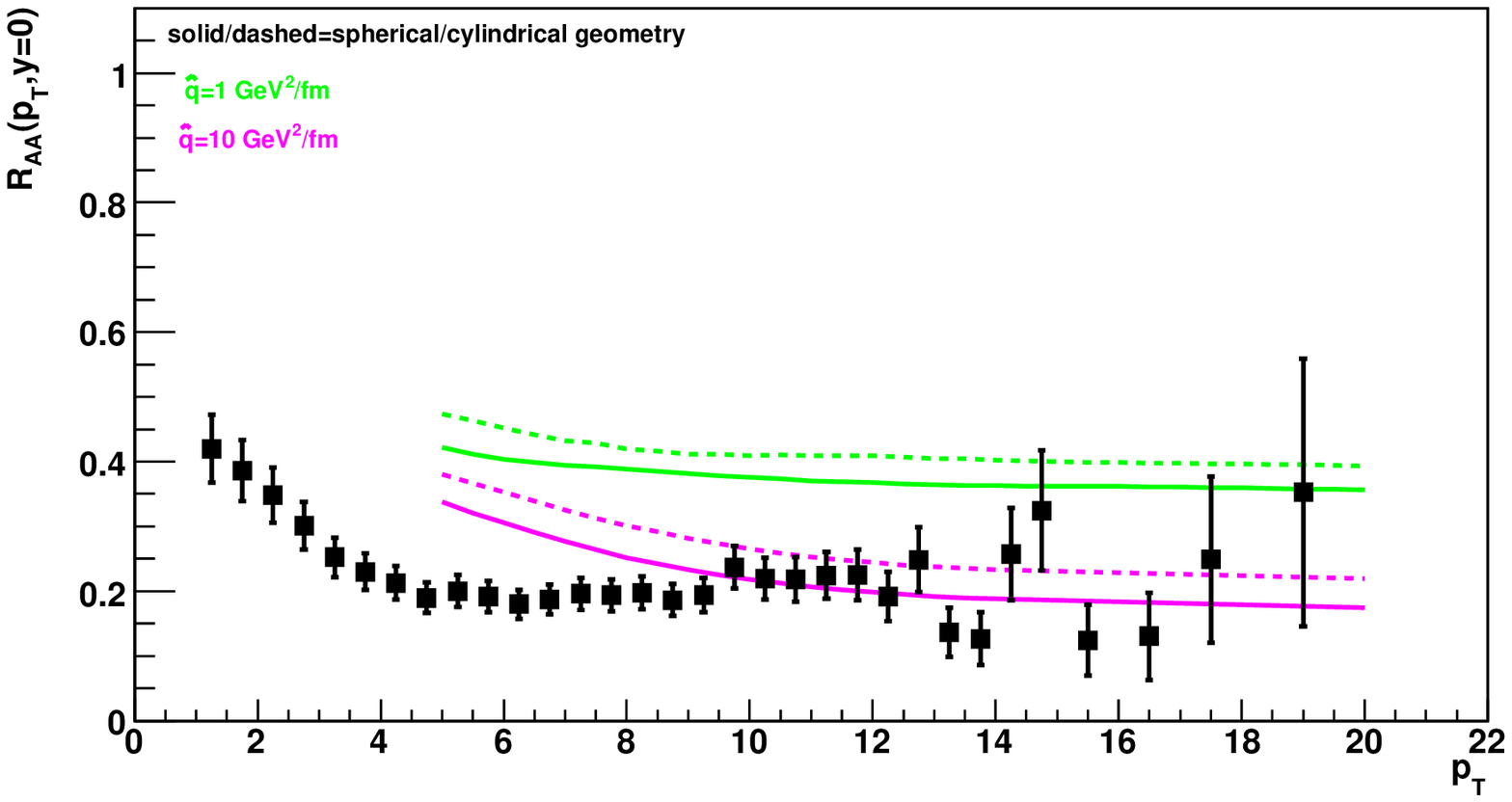}
\end{minipage}
\caption{Left: Nuclear modification factor $R_{AA}$ computed with the obtained medium-modified fragmentation functions for a fixed in-medium path-lenght of $L=6$ fm. Right: Same but computed with  more realistic geometries leading to a distribution of path-lenghts over which the suppression is averaged. In both cases, the experimental data is taken from \cite{Shimomura:2005en}.}
\label{fig321}
\end{figure}

In order to check  the validity of our formalism we first take a fixed length of  $L=6$ fm $\simeq R_{\rm Au}$ in the calculation of the MMFF. This simple geometry favors a value of $\hat q\sim$ 1 GeV$^2$/fm -- see Fig. \ref{fig321} -- in agreement with the findings in \cite{Salgado:2003gb}, where also a fixed length was taken. As it is known, the actual geometry used in the calculation affects the extracted value of the transport coefficient when the experimental data is fitted. It is not the goal of this paper to repeat fits of experimental data which have been extensively discussed in the literature in the past years. Here, we simply repeat the calculation with a geometry which takes into account the production point of the high-p$_T$ parton inside the extended medium. The procedure, performed for a cylindrical (constant profile) and for a spherical (constant density) nucleus, is as follows: i) we first sample a production point and a emission angle inside the medium by Monte Carlo; ii) the traversed length is then calculated and used to compute the cross section (\ref{eq:csec}) with medium effects; iii) the points i) and ii) are repeated to obtain an average cross section in nuclear collisions; iv) the ratio (\ref{eq:raa}) is done. The obtained result is plotted in Fig. \ref{fig321} where the favored value of $\hat q$ is larger by roughly one order of magnitude. Similar results were obtained in \cite{Eskola:2004cr}.

As a last comment, we have set the fragmentation scales equal to the fragmenting parton transverse momentum. Other common choice is to set the factorization scale to the final hadron transverse momentum, which is smaller by a factor $z$. In Fig. \ref{fig321} we show that both give similar results for the proton-proton case once the K-factors are adjusted to describe the data. However, the second choice would reduce the suppression, as it is evident from the Figs.  \ref{fig311} and \ref{fig311p}. Although the choice of scale is something arbitrary, this fact demonstrates the role of virtuality effects in jet quenching calculations. We find our implementation more natural in the picture of a parton branching process where the total amount of radiation is determined by the initial maximum virtuality. The latter is dictated by the perturbative hard cross section.

\section{Comparison with the approach based on quenching weights}
\label{sec3-3}

Medium modified fragmentation functions are usually computed by an energy shift of the corresponding vacuum fragmentation functions \cite{Wang:1996yh}:
\begin{equation}
D^{\rm med}(x,t)=  \int\frac{d\epsilon}{1-\epsilon}\, P(\epsilon) \, D^{\rm vac}\left(\frac{x}{1-\epsilon}
,t\right). 
\label{eq:FFQW}
\end{equation}
This model assumes that a highly energetic parton losses a fractional amount of energy $\epsilon$ while traveling through the medium and fragments with un-modified (vacuum) fragmentation functions once it is outside. Any modification of the virtuality dependence of the fragmentation is neglected and the probability distribution for the energy losses - quenching weights - has a discrete and a continuous part,
\begin{equation}
P(\epsilon)=p_0\delta(\epsilon)+p(\epsilon),
\end{equation}
given by
\begin{eqnarray}
p_0&=&\exp{\left[-\int d\omega \int d{\bf k}_\perp \frac{dI^{\rm med}}{d\omega
d{\bf k}_\perp} \right]}, \label{eq:QW0} \\
p(\epsilon)&=&p_0
\sum_{n=1}^\infty \prod_{i=1}^n \int d\omega_i \int d{\bf k}_{\perp i}
\,\frac{dI^{\rm med}}{d\omega_i
d{\bf k}_{\perp i}}\,\delta\left(\epsilon - \sum_{j=1}^n
\frac{\omega_i}{E}\right)\,.
\label{eq:QW}
\end{eqnarray}

In Appendix \ref{app1} we show that this procedure corresponds to the limiting case of our formalism for very large virtualities and considering only soft emissions. Here we show in Fig. \ref{fig331} the comparison of the results of our medium-modified evolution with those using the quenching weights \cite{Salgado:2003gb}. It can be seen that with increasing length of the evolution, the results of both approaches get closer and closer - specially for large values of $z$ most relevant for phenomenological applications -, in agreement with the discussion in Appendix \ref{app1}.

\FIGURE[h]{\epsfig{file=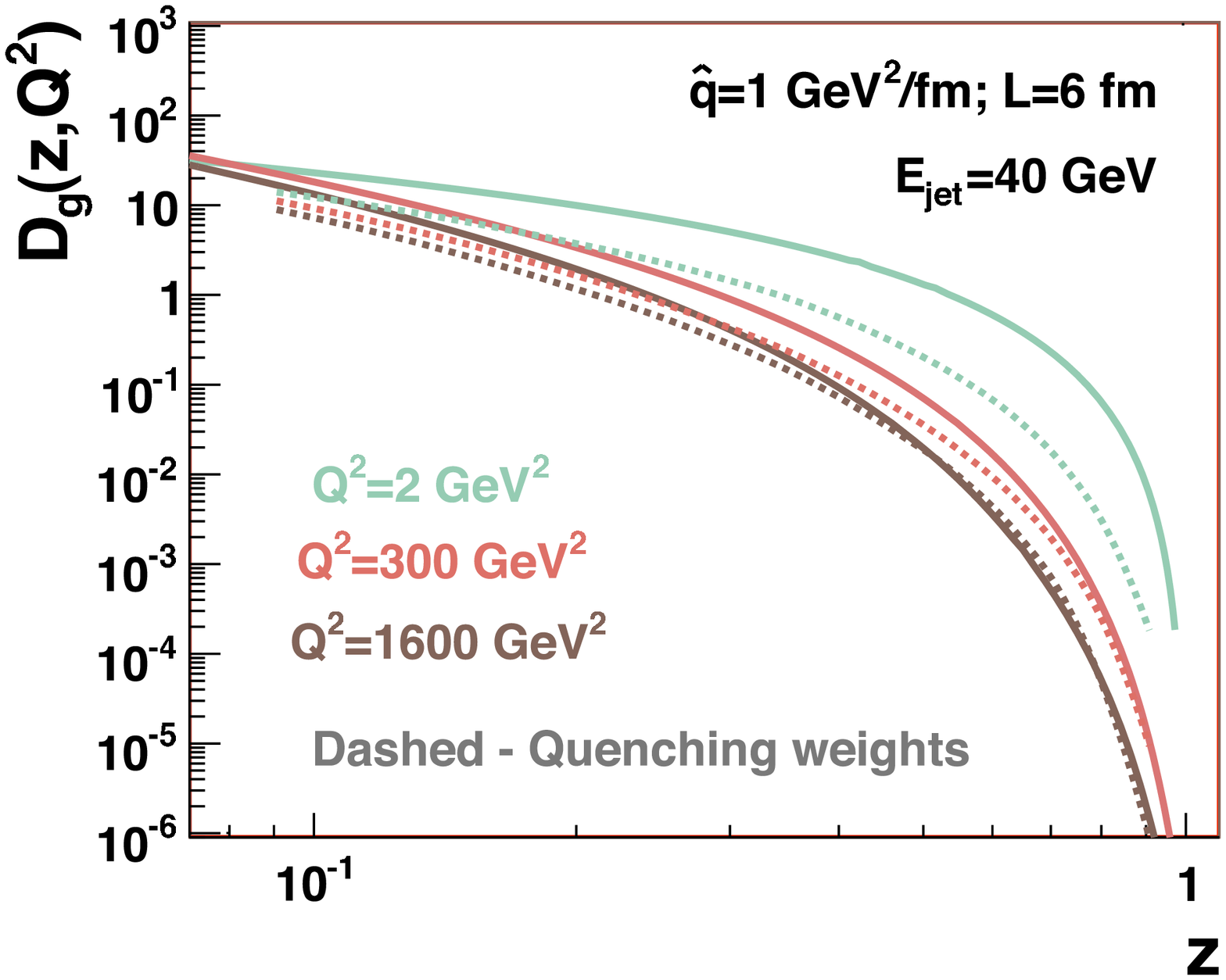,width=0.45\textwidth}\epsfig{file=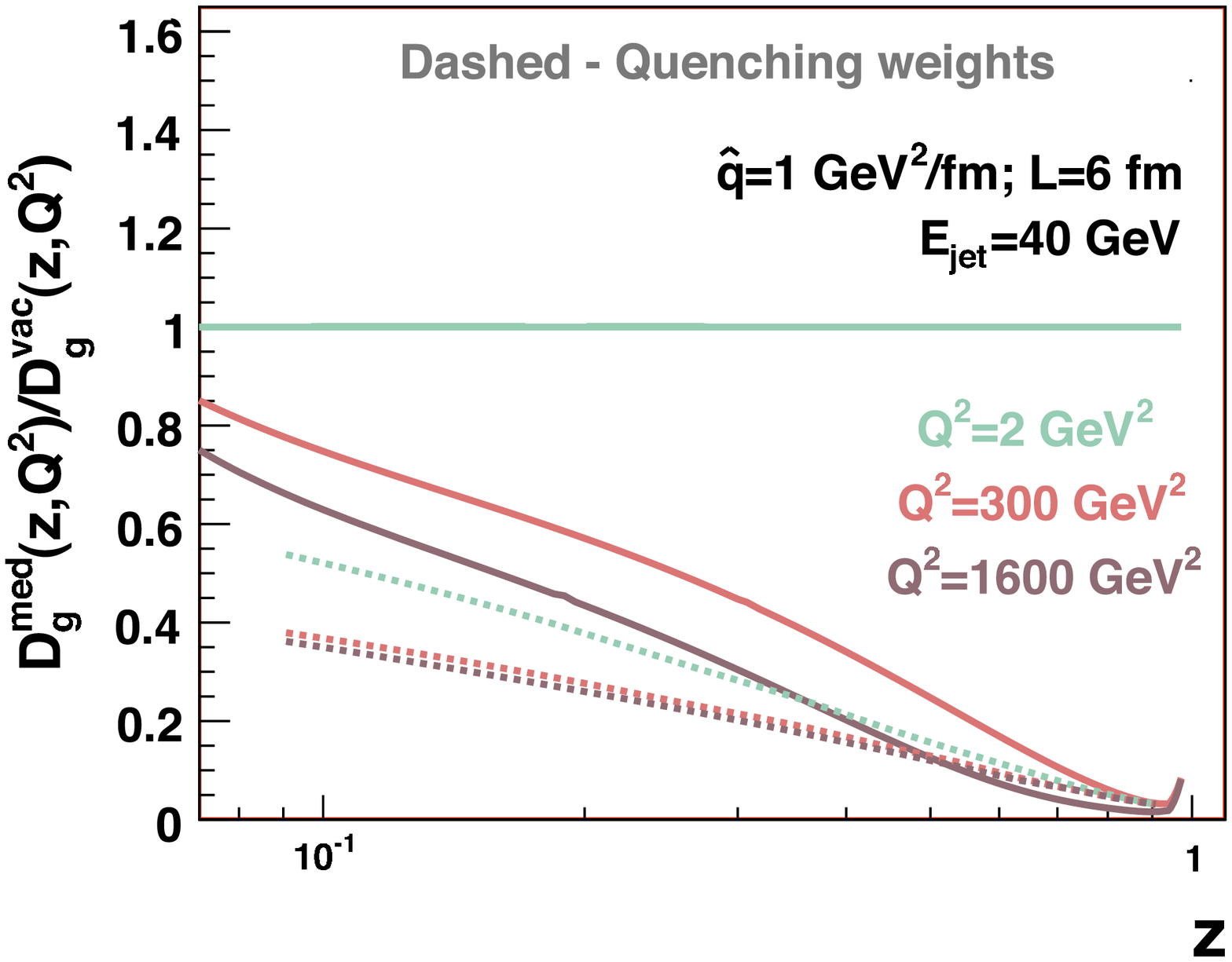,width=0.47\textwidth}
\caption{Left: Fragmentation function for gluons onto pions computed with our medium-modified evolution (solid lines) and through the standard convolution with the quenching weights (dashed lines), for $E_{jet}=40$ GeV, $\hat{q}=1$ GeV$^2$/fm, $L=6$ fm, and different $Q^2=2$, 300 and 1600 GeV$^2$. Right: Ratio of the fragmentation function for gluons onto pions in a medium with the same characteristic as for the plot on the left, over the fragmentation function in the vacuum, for the same values of $Q^2$.}
\label{fig331}}

\section{Conclusions}
\label{conclu}

We have presented a new implementation of medium effects in shower evolution which takes into account vacuum and medium-induced splittings on the same footing. We introduce a medium-modified splitting function into a DGLAP-like evolution. By doing so, our formalism improves previous ones, based on the Poisson approximation for the multiple gluon emission, by a consistent implementation of kinematic constrains on every individual splitting as well as on secondary emissions, which are included automatically . The explicit dependence of the virtuality in the evolution makes it possible to study its effects for the first time in the case of medium-induced gluon radiation. As expected, we obtain a softening of the fragmentation function. This softening increases with increasing virtuality, and with increasing medium length and density. In our implementation any effect on the initial non-perturbative condition is neglected since we do not expect it to change if large enough virtualities and parton energies are studied. The limited momentum range of applicability of the DGLAP fits \cite{Kniehl:2000fe} makes not possible to predict with reliability the {\it crossing point} between parton suppression at large-$z$ and parton enhancement at small-$z$, which also depends on the initial conditions chosen for the evolution \cite{Kniehl:2000fe,de Florian:2007hc}. This limitation will be circumvented in future works. 

One of the interests of our formalism is the simplicity to be included into a Monte Carlo code by implementing a medium-induced splitting probability $\Delta P(z,t)$ into the parton shower evolution routines. Work along this direction is in progress. In particular, it will allow to compute observables beyond single inclusive production, like two-and three-particle correlations, and compare the results with other approaches \cite{moredif}. It will also allow to study the possible interplay between the finite formation time of the partons radiated in subsequent emissions and the length of the medium, which has been ignored here as it is ignored in existing formalisms of energy loss \cite{Salgado:2003gb,quenwei,Gyulassy:2001nm, Jeon:2003gi}.

Finally, we have provided an independent check on the quality of the Poisson approximation for the calculation of inclusive particle production at high $p_T$. We have presented analytical and numerical checks of this approximation, and shown the agreement between both approaches for large values of the fraction of momentum. These are precisely the relevant values for the calculation of the inclusive particle suppression in heavy-ion collisions. This provides a connection with all the previous phenomenology and further supports the usual assumptions.

\acknowledgments
We thank N. Bianchi, M. Cacciari,
G. Corcella, D. D'Enterria, P. Di Nezza, D. de Florian, A. Morsch, A. H. Mueller and J.-W. Qiu
for useful discussions.
Special thanks are due to N. Borghini, F. Krauss
and
U. A. Wiedemann who participated in an early stage of this work. We also thank Centro de Supercomputaci\'on de Galicia for computer resources.
NA is supported by Ministerio de Educaci\'on
y Ciencia of Spain under a Ram\'on y Cajal contract; LC is supported by MEC under grant AP2005-3271;  CAS is supported by the 6th Framework Programme of the European Community under the
Marie Curie contract MEIF-CT-2005-024624;
NA, LC and CAS are supported by
CICYT of Spain under project FPA2005-01963 and by Xunta de Galicia
(Conseller\'{\i}a de Educaci\'on). WCX thanks Departamento de F\'{\i}sica de Part\'{\i}culas of the Universidade de Santiago de Compostela for warm hospitality during stays when part of this work was performed, and acknowledeges financial support from NSFC
(Grant No.10575044 and Key Grant No.10635020).

\appendix
\section{From the Sudakov form factors to the quenching weights}
\label{app1}
In this Appendix we show how the quenching weights
\cite{Salgado:2003gb,quenwei} appear when considering the DGLAP evolution of
the fragmentation functions. Ignoring for simplicity all parton and particles
labels, DGLAP evolution can be written as
\begin{equation}
D(x,t)=\Delta(t)D(x,t_0)+\Delta(t)\int_{t_0}^t \frac{dt_1}{t_1}
\frac{1}{\Delta(t_1)} \int \frac{dz}{z} \, P(z)
D\left(\frac{x}{z},t_1\right).
\label{eq-a1}
\end{equation}
The iterative solution to this equation reads
\begin{eqnarray}
D(x,t)&=& \Delta(t)D(x,t_0)+ \Delta(t)\sum_{n=1}^\infty \int_{t_0}^t
\frac{dt_1}{t_1} \int_{t_0}^{t_1} \frac{dt_2}{t_2}\cdots \int_{t_0}^{t_{n-1}}
 \frac{dt_n}{t_n} \int \frac{dz_1}{z_1} \int \frac{dz_2}{z_2} \cdots
\int \frac{dz_n}{z_n}\nonumber \\
&\times& P(z_1) P(z_2) \cdots P(z_n) D\left(\frac{x}{z_1z_2\cdots z_n},t_0
\right)\nonumber \\
&=& \Delta(t)D(x,t_0)+ \Delta(t)\int \frac{d\epsilon}{1-\epsilon}
\sum_{n=1}^\infty \int_{t_0}^t
\frac{dt_1}{t_1} \int_{t_0}^{t_1} \frac{dt_2}{t_2}\cdots \int_{t_0}^{t_{n-1}}
\frac{dt_n}{t_n} \prod_{i=1}^n\int dz_i \,P(z_i)
\nonumber \\
&\times& \delta\left(z_1z_2\cdots z_n-[1-\epsilon]\right)
D\left(\frac{x}{1-\epsilon},t_0
\right).
\label{eq-a2}
\end{eqnarray}

Now we consider $x_j=1-z_j\ll1$, $j=1,2,\dots,n$ i.e. the successive emissions
to take a small fraction of the parent energy-momentum, which results in
\begin{equation}
\delta\left(z_1z_2\cdots z_n-[1-\epsilon]\right)\simeq
\delta\left(\epsilon-\sum_{j=1}^n x_j\right)
\label{eq-a3}
\end{equation}
and
\begin{eqnarray}
D(x,t)&\simeq&
\Delta(t)D(x,t_0)+ \Delta(t)\int \frac{d\epsilon}{1-\epsilon}
\sum_{n=1}^\infty \frac{1}{n!} \prod_{i=1}^n \int_{t_0}^t
\frac{dt_i}{t_i}
\int dz_i\, P(z_i)
\nonumber \\
&\times& \delta\left(\epsilon-\sum_{j=1}^n x_j\right)
D\left(\frac{x}{1-\epsilon},t_0
\right).
\label{eq-a4}
\end{eqnarray}

Equation (\ref{eq-a4}) holds both for emissions in the vacuum (with
$\Delta^{\rm vac}(t)$, $P^{\rm vac}(z)$ and $D^{\rm vac}(x,t)$) and in the medium (with
$\Delta^{\rm med}(t)$, $\Delta P(z)$ and
$D^{\rm med}(x,t)$). Taking into account that in our
approach, due to the property (\ref{2.4}), we have
\begin{equation}
P(z)=P^{\rm vac}(z)+\Delta P(z),\ \ \Delta(t)=\Delta^{\rm vac}(t)\Delta^{\rm med}(t),
\label{eq-a5}
\end{equation}
we get
\begin{eqnarray}
D(x,t)&\simeq&
\Delta^{\rm med}(t)D^{\rm vac}(x,t)+ \Delta^{\rm med}(t)\int \frac{d\epsilon}{1-\epsilon}
\sum_{n=1}^\infty \frac{1}{n!} \prod_{i=1}^n \int_{t_0}^t
\frac{dt_i}{t_i}
\int dz_i\, \Delta P(z_i)
\nonumber \\
&\times& \delta\left(\epsilon-\sum_{j=1}^n x_j\right)
D^{\rm vac}\left(\frac{x}{1-\epsilon},t
\right),
\label{eq-a6}
\end{eqnarray}
where the sum and product run now over medium-induced emissions and we have
used the fact that
\begin{eqnarray}
&&\int \frac{d\epsilon}{1-\epsilon} \delta\left(\epsilon-\sum_{\rm vac}
x_i-\sum_{\rm med} x_j\right)
D\left(\frac{x}{1-\epsilon},t_0
\right)
\label{eq-a7} \\
&\simeq& \int \frac{d\epsilon^\prime}{1-\epsilon^\prime}
\delta\left(\epsilon^\prime
-\sum_{\rm med} x_j\right)
\int \frac{d\epsilon^{\prime\prime}}{1-\epsilon^{\prime\prime}}
\delta\left(\epsilon^{\prime\prime}-\sum_{\rm vac}
x_i\right)
D\left(\frac{x}{(1-\epsilon^\prime)(1-\epsilon^{\prime\prime})},t_0
\right),
\nonumber
\end{eqnarray}
with $(1-\epsilon^\prime-\epsilon^{\prime\prime})\simeq
(1-\epsilon^\prime)(1-\epsilon^{\prime\prime})$
for $\epsilon^\prime, \epsilon^{\prime\prime}\ll 1$.

Finally, we ignore virtuality (as done in all previous approaches to radiative
energy loss except in \cite{Wang:2001if}) and set in the integrals over $t_i$
and $z_i$ the kinematical limits for gluon emission. Then we can identify,
using (\ref{scalvar}),
 (\ref{vacsplit}) and (\ref{medsplit}), the discrete and
continuous parts of the quenching weights \cite{Salgado:2003gb,quenwei}
\begin{eqnarray}
p_0&=&\exp{\left[-\int d\omega \int d{\bf k}_\perp \frac{dI^{\rm med}}{d\omega
d{\bf k}_\perp} \right]}, \label{eq-a8} \\
p(\epsilon)&=&p_0
\sum_{n=1}^\infty \prod_{i=1}^n \int d\omega_i \int d{\bf k}_{\perp i}
\,\frac{dI^{\rm med}}{d\omega_i
d{\bf k}_{\perp i}}\,\delta\left(\epsilon - \sum_{j=1}^n
\frac{\omega_i}{E}\right)
\label{eq-a9}
\end{eqnarray}
to get the standard
expression for the fragmentation functions in a medium through
the quenching weights \cite{Wang:1996yh}:
\begin{equation}
D(x,t)\simeq p_0 \,D^{\rm vac}(x,t) + \int
\frac{d\epsilon}{1-\epsilon}\, p(\epsilon) \, D^{\rm vac}\left(\frac{x}{1-\epsilon}
,t\right). 
\label{eq-a10}
\end{equation}

Summarizing, DGLAP evolution of fragmentation functions in a medium
corresponds to the usual picture of radiative energy loss through
quenching weights when virtualities are ignored, and only soft emissions are
considered. This latter consideration
lies at the root of all existing formalisms for radiative
energy loss. In this correspondence, the discrete part of the
quenching weights maps onto the Sudakov form factor, (\ref{eq-a8}).


\end{document}